\documentclass[twocolumn,twoside,slac_two]{revtex4}
\pdfoutput=1
\usepackage{graphicx}
\usepackage{fancyhdr}
\begin{document}

\title{\boldmath CP Violation in Hadronic $\tau$ Decays}

\author{K. Kiers}
\affiliation{Physics Department, Taylor University, 
236 West Reade Ave., Upland, IN 46989, USA}

\begin{abstract}
We examine CP violation in the $\Delta S=0$ decays 
$\tau\to\omega \pi \nu_\tau$ and $\tau\to a_1\pi\nu_\tau$
and the $\Delta S=1$ decay $\tau\to K
\pi\pi\nu_\tau$.  We assume that the new physics is a charged
Higgs.  We show that sizeable
CP-violating effects are possible in $\tau\to a_1\pi\nu_\tau$ 
(polarization-dependent rate asymmetry) and
$\tau\to\omega\pi\nu_\tau$ (triple-product asymmetry).  The $\Delta
S=1$ decay $\tau\to K \pi\pi\nu_\tau$ can proceed via several
resonances.  We construct two modified rate asymmetries and a triple
product asymmetry for this decay and discuss the potential 
sensitivities of these asymmetries.
\end{abstract}

\maketitle

\thispagestyle{fancy}


\section{Introduction}
\label{sec:intro}
In the Standard Model (SM) of particle physics, CP violation is due to
a complex parameter in the Cabibbo-Kobayashi-Maskawa (CKM) matrix.  
Extensions of the SM typically include new sources of CP violation.  
To understand the origin of CP violation it is important to
investigate as many systems as possible.

One area that deserves further investigation is the $\tau$ system.  
We consider a few hadronic decay modes of the $\tau$.  In the SM, 
there is virtually no CP violation in these decay modes.  Thus, any 
observation of CP violation for these decays would be a clear 
indication of New Physics (NP).

\begin{figure}[t]
\centering
\includegraphics[width=40mm]{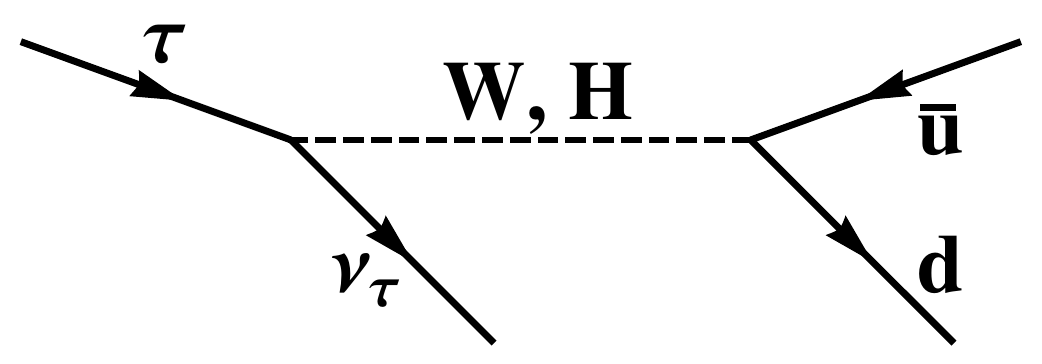}
\caption{Quark-level Feynman diagram for the $\Delta S=0$
decays showing the $W$ and $H$ contributions.} 
\label{fig:feyndiag}
\end{figure}

CP-odd observables require at least two contributing amplitudes 
in order to be non-zero.  In addition to the usual W-exchange 
amplitude that is present in the SM, we assume that there is 
also a contribution that arises due to the exchange of a 
charged Higgs boson -- see Fig.~\ref{fig:feyndiag}.

Many extensions of the SM include extra Higgs bosons.  
The charged Higgs couplings to light quarks in these 
models are often proportional to the quark masses, and 
are thus very small.  We investigate $\tau$ decays that contain 
light quarks in the final 
state~\cite{CPVhadronictau,CPVKpipi}.\footnote{
Portions of the current work (excerpts of text,
equations, etc.) are reprinted with permission from
Alakabha Datta, Ken Kiers, David London, Patrick J. O'Donnell 
and Alejandro Szynkman, Physical Review D {\bf 75}, 074007 (2007) 
[Erratum-ibid. D {\bf 76}, 079902 (2007)].  Copyright (2007)
by the American Physical Society.
The reader is referred to 
Refs.~\cite{CPVhadronictau,CPVKpipi} for more details.}
If CP violation 
is to be large in these decays, the charged Higgs couplings 
must be large.  Thus, these decays probe ``non-standard'' 
NP CP violation.

\section{CP Asymmetries}
\label{sec:cpasyms}

Suppose two amplitudes contribute to a particular 
process.  Then the total amplitude may be written as 
\begin{eqnarray}
    {\cal A} = A_1 + A_2 e^{i\phi} e^{i\delta} ~,
\end{eqnarray}
where $\phi$ and $\delta$ are the relative weak (CP-odd) and strong
(CP-even) phases, respectively. 

The full rate for the process is obtained by 
calculating $|{\cal A}|^2$, summing/averaging over spins, and 
integrating over phase space.  The regular 
rate asymmetry for a particular decay mode is 
proportional to the difference between the rate 
for the process and that for its associated 
anti-process.  The result is proportional to 
\begin{eqnarray}
  \sin\phi\sin\delta
\end{eqnarray}
and thus requires both a weak phase difference ($\phi$) 
and a strong phase difference ($\delta$) between 
the contributing amplitudes.

The rate asymmetry can be altered in various ways.  
For example, if some of the spins are measured, 
one does not sum over them.  One can also integrate 
asymmetrically over phase space (or use a more 
general weighting function) to isolate certain 
terms in the differential width.  In some cases 
this leads to asymmetries that are proportional 
to $\sin\phi\sin\delta$, similar to the regular rate asymmetry.  
One particular class of asymmetries involves triple 
products, which have the form 
$\vec{v}_1\cdot\left(\vec{v}_2\times\vec{v}_3\right)$,
where $\vec{v}_1$, $\vec{v}_2$ and $\vec{v}_3$
are momenta and/or spins.  CP asymmetries constructed 
from triple products have the form
\begin{eqnarray}
  \sin\phi\cos\delta
\end{eqnarray}
and thus require a weak phase difference, but 
not a strong phase difference.

\section{\boldmath Decays with $\Delta S=0$: $\tau\to
\omega\pi\nu_\tau$ and $\tau\to a_1\pi\nu_\tau$}

\subsection{Form Factors}

Consider the decay $\tau\to V\pi\nu_\tau$,
where $V$ represents a vector or 
axial vector meson.  The general structure for the 
SM hadronic current for this decay is given by~\cite{decker87,deckermirkes},
\begin{eqnarray}
    J^{\mu} & = & \langle V(q_1)\pi(q_2)|\bar{d}\gamma^\mu
    \left(1-\gamma^5\right)u|0\rangle \nonumber\\
    & = & F_1(Q^2)\left(
    Q^2\epsilon_1^\mu-\epsilon_1\cdot q_2 Q^\mu\right)\nonumber\\
    & & \!\!\!\!\!\!\!\!\!\!\! +F_2(Q^2)\,\epsilon_1\cdot q_2
         \left(
         q_1^\mu-q_2^\mu-Q^\mu\frac{Q\cdot(q_1-q_2)}{Q^2}\right)~~~~
         \nonumber\\
    & & \!\!\!\!\!\!\!\!\!\!\! +iF_3(Q^2)\,
         \varepsilon^{\mu\alpha\beta\gamma}
             \epsilon_{1\alpha}q_{1\beta}q_{2\gamma}
       +F_4(Q^2)\,\epsilon_1\cdot q_2 Q^\mu ,
\label{eq:Wamp}
\end{eqnarray}
where $Q^\mu\equiv (q_1+q_2)^\mu$ and 
where $\epsilon_1$ denotes the
polarization tensor of the $V$.  Experimentally, 
the form factor $F_3$ dominates for $V =\omega$ and $F_1$ 
and/or $F_2$ are expected to dominate for $V = a_1$.  
The scalar form factor, $F_4$, is expected to be 
very small within the SM.

The NP charged Higgs effect may be parameterized 
in terms of  new scalar and pseudo-scalar 
quark-level operators.  The resulting current is given by,
\begin{eqnarray}
  J_{\textrm{\scriptsize{Higgs}}} 
  & = & \langle
  V(q_1)\pi(q_2)|\bar{d}\left(a+b\gamma^5\right)u|0\rangle \nonumber\\
  &=&  
  \cases{
    bf_{H} \, \epsilon_1 \cdot q_2, & $V=\omega$, \cr
    af_{H} \, \epsilon_1 \cdot q_2, & $V=a_1$,}
\label{eq:Hamp}
\end{eqnarray}
where $f_H$ is a form factor for the quark-level 
operators and $a$ and $b$ arise from the coupling of 
the charged Higgs boson to the quarks and leptons.  
The charged Higgs effect can be incorporated 
into  the expression for the hadronic current 
[Eq.~(\ref{eq:Wamp})] by the replacement
$F_4(Q^2)\to \widetilde{F}_4(Q^2)$, where
\begin{eqnarray}
\widetilde{F}_4(Q^2)=
\cases{
F_4(Q^2)+ ({bf_H}/{m_\tau}), & $V=\omega$, \cr
F_4(Q^2)+ ({af_H}/{m_\tau}), & $V=a_1$.}
\end{eqnarray}
The parameters $a$ and $b$ can be complex; that is, 
they can contain a (CP-odd) weak phase.  The 
various form factors are potential sources of strong phases.

\subsection{\boldmath Results for $\Delta S=0$}

\subsubsection{Rate Asymmetry}

The regular rate asymmetry is proportional to 
$|F_4 f_H b|\sin\left(\delta_4-\delta_H\right)
\sin\left(\phi_b\right)$ 
for the $V =\omega$ case (integrated over phase space), where 
the $\delta$'s are the strong phases associated with the form 
factors and $\phi_b$ is the (CP-odd) phase of the complex Higgs 
coupling $b$.  An analogous expression (with $b$ replaced 
by $a$) holds for the case $V = a_1$.  Since $F_4$ is expected 
to be very small, we conclude that the rate asymmetry 
is unlikely to be measureable, even in the presence of NP.

\subsubsection{Polarization-dependent Rate Asymmetry}

Weighting the differential width by $\cos\beta$ while performing 
the integration over phase space ($\beta$ is a particular 
kinematical angle -- see Ref.~\cite{CPVhadronictau}) extracts terms containing 
the combinations $F_1 f_H^*$ and $F_2 f_H^*$.  Such terms could be 
non-vanishing for the decay $\tau\to a_1\pi\nu_\tau$
(since $F_1$ and/or $F_2$ are 
expected to be the dominant SM form factors in this case).  
Constructing a CP asymmetry from this quantity yields an 
expression that depends on the polarization of the $\tau$
and that contains pieces such as
$|F_1f_Ha|\sin\left(\delta_1-\delta_H\right)
\sin\left(\phi_a\right)$.
This asymmetry requires a 
strong phase difference between $F_1$ (or $F_2$) and $f_H$.  
Numerical estimates (see Ref.~\cite{CPVhadronictau} 
for details) indicate 
that asymmetries of order 15\% (7.5\%) could be possible, 
given the uncertainties in the experimental measurement 
of the branching ratio.  The first estimate assumes that only $F_2$ 
contributes to the SM amplitude; the second that only $F_1$ does.

\subsubsection{Triple-product Rate Asymmetry}

A triple product (TP) can be constructed using the 
polarization tensor of the vector meson.  The TP 
contains the combination of form factors $F_3 f_H^*$, and 
could thus be non-zero for the decay 
$\tau\to\omega\pi\nu_\tau$
(for which $F_3$ 
dominates the SM hadronic current).  Constructing a 
CP asymmetry from the TP yields an expression that 
contains 
$|F_3 f_H b|\cos\left(\delta_3-\delta_H\right)\sin\left(\phi_b\right)$; 
thus this CP asymmetry 
does not require the presence of a relative strong 
phase between the SM and NP amplitudes.  A numerical 
estimate performed in Ref.~\cite{CPVhadronictau} 
indicates that the TP 
asymmetry could be as large as 30\% multiplied by 
$\left(\vec{\epsilon}_1\cdot\vec{n}_1\right)
\left(\vec{\epsilon}_1\cdot\vec{n}_2\right)$,
where $\vec{n}_1$ and $\vec{n}_2$ are particular direction vectors 
in the hadronic rest frame and $\vec{\epsilon}_1$ is the 
polarization vector of the $\omega$.

\section{\boldmath A Decay with $\Delta S=1$: $\tau\to
  K\pi\pi\nu_\tau$}

\subsection{Overview and Preliminary Results}

The previous approach can be generalized to the 
decay $\tau\to K\pi\pi\nu_\tau$.  The quark-level process 
is the same as that shown in Fig.~\ref{fig:feyndiag}, but 
with $d\to s$.  The hadronic current is given by~\cite{CPVKpipi,KM1992,dmsw},
\begin{eqnarray}
   J^\mu & \equiv & \langle K^-(p_1) \pi^-(p_2) \pi^+(p_3)| \bar{s}\gamma^\mu
     (1-\gamma^5) u | 0\rangle \nonumber \\
     &=& \left[F_1(s_1,s_2,Q^2) (p_1-p_3)_\nu \right. \nonumber\\
     & & +F_2(s_1,s_2,Q^2) (p_2-p_3)_\nu \left.\right]T^{\mu\nu} \nonumber \\
     && + i F_3(s_1,s_2,Q^2) \epsilon^{\mu\nu\rho\sigma}
         p_{1\nu}p_{2\rho}p_{3\sigma} \nonumber \\
     && + F_4(s_1,s_2,Q^2) Q^\mu\, ,
\label{eq:jmu_sm}
\end{eqnarray}
where $Q^\mu=(p_1+p_2+p_3)^\mu$, $T^{\mu\nu}=g^{\mu\nu}-Q^\mu
Q^\nu/Q^2$, $s_1=(p_2+p_3)^2$ and $s_2=(p_1+p_3)^2$.
Several decay chains contribute 
to the form factors within the SM.  
The dominant form factors $F_1$ and $F_2$ have
been studied experimentally by CLEO~\cite{CLEOKpipi}.
These form factors
receive contributions due to 
$\tau\to K_1^-\nu_\tau\to K^*\pi^-\nu_\tau\to K^-\pi^-\pi^+\nu_\tau$
and
$\tau\to K_1^-\nu_\tau\to \rho K^-\nu_\tau\to K^-\pi^-\pi^+\nu_\tau$,
respectively.  The subdominant processes
$\tau\to K^*\nu_\tau\to K^*\pi^-\nu_\tau\to K^-\pi^-\pi^+\nu_\tau$
and 
$\tau\to K^*\nu_\tau\to \rho K^-\nu_\tau\to K^-\pi^-\pi^+\nu_\tau$
are also possible, and could contribute to $F_3$.  
The scalar form factor, $F_4$ , is expected 
to be small within the SM.  The NP 
charged Higgs  contribution may be taken 
into account by the replacement,
\begin{eqnarray}
   F_4 & \rightarrow & \widetilde{F}_4=F_4
     + \frac{f_H}{m_\tau}\left(\eta_{RL}-\eta_{LL}\right) \, ,
\label{eq:f4_mod}
\end{eqnarray}
where $f_H$ is the pseudoscalar form factor 
and $\eta_{RL}$ and $\eta_{LL}$ are (possibly complex) 
NP parameters defined somewhat similarly to $a$ and $b$ in 
the $\Delta S = 0$ case.

We again consider several possible CP asymmetries.  
Each asymmetry is proportional to 
$|f_H|\mbox{Im}\left(\eta_{RL}-\eta_{LL}\right)$
(integrated over phase space).  The regular 
rate asymmetry contains the combination 
$|F_4 f_H|\mbox{Im}\left(\eta_{RL}-\eta_{LL}\right)$ 
and is thus expected 
to be very small.  It is also possible to 
construct other asymmetries by employing 
various weighting functions when performing 
the integration over phase space (see, for example,
Refs.~\cite{KM1992,kilian}).  In this 
manner we have constructed two modified rate 
asymmetries, as well as one TP asymmetry.  
The modified rate asymmetries contain the 
SM hadronic currents $F_1$ and $F_2$ and they 
require a non-zero relative strong phase 
between the SM and NP amplitudes in order 
to be non-zero.  The TP asymmetry depends 
on $F_3$ and does not require a relative strong 
phase between the interfering amplitudes.

A preliminary numerical analysis of the modified 
and TP asymmetries indicates that some amount of 
cancellation tends to occur as one performs the 
integrations over phase space.  These cancellations 
would tend to make the asymmetries quite small.  
Larger asymmetries are possible if one includes 
extra weighting factors to offset the cancellations.
Alternatively, experimentalists could perform fits
to differential asymmetries (such as $d a_{CP}/dQ$, 
for example).  
Our initial study indicates that asymmetries of 
up to the order of a percent might be possible if the only assumption
regarding the NP scalar
contribution is that it is ``hidden'' in the current 
experimental uncertainty in the branching ratio.
Incorporating the CLEO bound on the scalar coupling
coming from $\tau\to K\pi\nu_\tau$~\cite{CLEObound}
and making a reasonable estimate for the scalar
form factor, we find that the CP asymmetries are likely to be 
smaller than this.
Further analysis and refinement
of the numerical estimates
will be provided in Ref.~\cite{CPVKpipi}.

\bigskip
\noindent
[{\bf Note added:}
As noted in Ref.~\cite{CPVKpipi} (and contrary to the statement
made above),
the CLEO bound does not actually place a direct constraint on the NP
coupling considered here.  In the notation of Ref.~\cite{CPVKpipi},
the CLEO experiment placed a bound on $\eta_S$, while
$\tau\to K \pi \pi \nu_\tau$ probes $\eta_P$.
For more details, please see Ref.~\cite{CPVKpipi}.]

\bigskip

\section{Conclusions}

We have considered CP violation in certain $\Delta S = 0$ 
and $\Delta S = 1$  $\tau$ decays.  In both cases CP violation 
requires the interference of (at least) two 
amplitudes that have a differing weak phase.  
One amplitude is provided by the usual SM W-exchange 
diagram.  The other amplitude is assumed to be 
due to a NP Higgs-exchange diagram.  While the 
regular rate asymmetries are expected to be 
very small in these decays, larger asymmetries 
can be obtained by forming triple products or 
by considering other modifications to the 
usual rate asymmetries.  

\bigskip

\begin{acknowledgments}
It is a pleasure to thank those with whom I have collaborated on this
work -- A. Datta, K. Little, D. London, M. Nagashima, P.J. O'Donnell
and A. Szynkman.  I also thank the conference organizers for an
excellent conference.  This work was supported in part by the U.S. National
Science Foundation under Grants No. PHY-0301964 and No. PHY-0601103.
\end{acknowledgments}


\end{document}